\documentclass[journal]{IEEEtran}

\usepackage{amssymb,amsthm,amsmath}
\usepackage{graphicx}                       
\usepackage{epstopdf}
\usepackage[usenames,dvipsnames]{xcolor}    
\usepackage{siunitx}                        
\usepackage{subfigure}                      
\usepackage{fancyhdr}                       
\usepackage{balance}                        
\usepackage[hidelinks]{hyperref}            

\graphicspath{{../}{./figures/}}
\UseRawInputEncoding

\newcommand{\tx}{\mathrm}

\usepackage{booktabs}
\usepackage{threeparttable}
\usepackage{multirow}
\usepackage{url}

\usepackage{enumitem}

\usepackage{booktabs}
\usepackage{stfloats}
 \usepackage{cite}

\usepackage{xcolor,soul,framed} 
\colorlet{shadecolor}{yellow}

\begin{document}
\title{Adaptive Leading Cruise Control in Mixed Traffic\\ Considering Human Behavioral Diversity}
%
%

\author{Qun Wang,
        Haoxuan Dong,
        Fei Ju,
        Weichao Zhuang,
        Chen Lv,
        Liangmo Wang,
        and Ziyou Song 
\thanks{Qun Wang is with the Department of Mechanical Engineering, National University of Singapore, Singapore 117575, Singapore, and also with the School of Mechanical Engineering, Nanjing University of Science and Technology, Nanjing 210094, China (e-mail: wangqun@u.nus.edu).}
\thanks{Haoxuan Dong and Weichao Zhuang are with the School of Mechanical Engineering, Southeast University, Nanjing 211189 , China (e-mail: donghaox@foxmail.com; wezhuang@seu.edu.cn).}
\thanks{Fei Ju and Liangmo Wang are with the School of Mechanical Engineering, Nanjing University of Science and Technology, Nanjing 210094, China (e-mail:jufei@njust.edu.cn; liangmo@njust.edu.cn).}
\thanks{Chen Lv is with the School of Mechanical and Aerospace Engineering, Nanyang Technological University, Singapore 639798 (e-mail: lyuchen@ntu.edu.sg).}
\thanks{Ziyou Song is with the Department of Mechanical Engineering, National University of Singapore, Singapore 117575, Singapore (e-mail: ziyou@nus.edu.sg).}
\thanks{Corresponding authors: Weichao Zhuang; Ziyou Song.}}
%
%

\markboth{}%
{Shell \MakeLowercase{\textit{et al.}}: Bare Demo of IEEEtran.cls for IEEE Journals}



\maketitle 

\begin{abstract}
This paper presents an adaptive leading cruise control strategy for the connected and automated vehicle (CAV) and first considers its impact on the following human-driven vehicle (HDV) with diverse driving characteristics in the unified optimization framework for improved holistic energy efficiency. 
The car-following behaviors of HDV are statistically calibrated using the Next Generation Simulation dataset.
In a typical single-lane car-following scenario where CAVs and HDVs share the road, the longitudinal speed control of CAVs can substantially reduce the energy consumption of the following HDV by avoiding unnecessary acceleration and braking. Moreover, apart from the objectives including car-following safety and traffic efficiency, the energy efficiencies of both CAV and HDV are incorporated into the reward function of reinforcement learning. The specific driving pattern of the following HDV is learned in real-time from historical speed information to predict its acceleration and power consumption in the optimization horizon. A comprehensive simulation is conducted to statistically verify the positive impacts of CAV on the holistic energy efficiency of the mixed traffic flow with uncertain and diverse human driving behaviors. Simulation results indicate that the holistic energy efficiency is improved by \SI{4.38}{\%} on average.
\end{abstract}


%
\IEEEpeerreviewmaketitle

\section{Introduction}
%
%
%
%
\IEEEPARstart{T}{he} transportation sector, which consumes 25\% of global energy resources, is one of the main sources of greenhouse gas emissions and air pollution \cite{mohsin2019integrated}. Extensive efforts have been made to improve vehicle efficiency and lower emissions of on-road vehicles in response to the increasingly stringent emission standards \cite{mahler2014optimal, guo2021supervisory, chen2019series}. As a crucial technology in saving energy consumed by vehicles, eco-driving has been extensively discussed \cite{barkenbus2010eco, barth2009energy, rakha2011eco, barth2011dynamic, dong2022predictive}, with the core idea of adjusting vehicle speed and maintaining an energy-efficient driving style \cite{jin2016power}. More recently, the development of connectivity and automation technologies has provided another promising opportunity to further cut down energy consumption through the deployment of connected and automated vehicles (CAVs). With the assistance of vehicle-to-anything (V2X) communications \cite{abboud2016interworking} and sensor fusion techniques \cite{garcia2017sensor}, CAVs can take advantage of the rich information to optimize their operations, such as vehicle acceleration \cite{wegener2021automated}, motor torque regulation \cite{xu2020parametric}, path planning \cite{yanumula2021optimal}, etc. 

Despite some promising results indicating that CAVs can save energy, the impacts of connectivity and autonomy on the traffic efficiency and energy performance of neighboring vehicles have not been extensively studied, while this type of study can provide insights for policymakers and incentives to further promote CAVs. For example, Joshua \textit{et al.} \cite{auld2018impact} analyzed the mobility and energetic impacts introduced by CAVs' deployment. Results demonstrate that the traffic flow is improved with the increasing travel demand and decreasing travel time. Fakhrmoosavi \textit{et al.} \cite{fakhrmoosavi2022stochastic} explored the influences of a mixed traffic fleet on several aspects from a network level, indicating that CAVs can enhance traffic safety, mobility, and emission reduction of the traffic system. Zhao \textit{et al.} \cite{zhao2018anticipating} assessed the impacts of CAVs under eight different testing scenarios with a travel demand model and simulation results indicate that the travel demand in Austin, Texas can increase by at least 20\%. 

Moreover, it is anticipated that CAVs and human-driven vehicles (HDVs) will co-exist on the same road in the near future \cite{stern2018dissipation, zheng2020smoothing}. Human drivers will still remain to be the majority who take charge of vehicle operations for a long period. Hence, it is imperative to develop eco-driving strategies for CAVs in the mixed traffic flow in which CAVs frequently interact with HDVs.  
Lu \textit{et al.} \cite{lu2019energy} proposed an energy-efficient adaptive cruise control model for electric CAVs in a mixed traffic flow. Simulations are performed in a mixed single-lane traffic flow with different market penetration rates of CAVs, indicating that the proposed method exhibits a superior performance in energy saving compared to other existing adaptive cruise control and cooperative adaptive cruise control methods.
Zhu \textit{et al.} \cite{Zhu_Song_Zhuang_Hofmann_Feng_2021} designed a novel model predictive control (MPC) method to enhance energy efficiency and keep driving safety for the CAV in a mixed traffic flow.
An integrated data-driven model of car-following is used in the MPC framework to predict the behaviors of HDVs. Simulation results validate its effectiveness in energy efficiency improvement and robustness.
Li \textit{et al.} \cite{li2021reinforcement} developed a cooperative controller for CAVs in a mixed traffic platoon based on multi-agent reinforcement learning. Compared with MPC, the proposed strategy performs better in dampening traffic oscillations and reducing energy consumption.
Ma \textit{et al.} \cite{ma2020energetic} investigated the energy-saving potentials of the following human-driven platoon enabled by eco-driving control of CAVs ahead. Especially, the influences of diverse characteristics of human behaviors are evaluated through extensive numerical analyses, which statistically show a positive influence of the proposed strategy on the subsequent platoon.

However, all the aforementioned studies only focus on the optimization of the CAV, while neglecting its impact on the following HDV. To achieve a higher holistic energy efficiency of the mixed traffic flow, this study first incorporates the HDV energy consumption in the optimization framework. 
In most existing studies, car-following models (e.g., optimal velocity model \cite{sugiyama1999optimal} and intelligent driver model (IDM) \cite{kesting2010enhanced}) are usually utilized to describe the behaviors of HDVs. 
In \cite{zhu2018modeling}, five representative microscopic car-following models were used to calibrate the behaviors of drivers in Shanghai, and the IDM outperformed the other models from the perspectives of accuracy and stability.
The parameters of these models indicating different driving styles are either set as constant \cite{zhao2018platoon}, or assumed to follow some predefined uniform distribution \cite{ma2020energetic}. However, in a dynamic traffic environment, the behaviors of HDVs are quite stochastic and do not follow deterministic patterns. It remains challenging to accurately predict the behaviors of HDVs, which are the necessary previews of most predictive control schemes.

More recently, model-free reinforcement learning (RL) algorithms has been widely applied in many areas such as autonomous driving \cite{wu2022toward}, \cite{aradi2020survey}, battery management \cite{zhuang2020comparison}, and eco-driving for electrified vehicles \cite{wang2022ecological}, \cite{lee2022energy}. One of the main advantages of model-free RL is that the agent can interact with the stochastic environment and try to maximize the accumulated reward in a learning manner. Instead of attempting to model the complicated environment with high stochasticity (e.g., uncertain human driving behaviors in this study), model-free methods directly improves system performance based on the explored samples \cite{dong2020deep}.

Motivated by the discussion above, this study aims to design an adaptive leading cruise control strategy to reduce the holistic energy consumption of both CAV and HDV by considering the diverse human driving behaviors in a reinforcement learning framework. The contributions and novelties of this study are summarized as follows:
\begin{enumerate}
    \item In addition to car-following safety and traffic efficiency, the dynamics of both CAV and the following HDV are considered in the optimization framework for improved holistic energy efficiency.
    \item HDVs are calibrated into a joint distribution using the IDM based on the field-collected Next Generation Simulation (NGSIM) dataset to cover a wide range of stochastic and realistic driving behaviors.
    \item The influences of diverse driving behaviors on the improvement of energy efficiency using the proposed control algorithm are quantitatively analyzed.
\end{enumerate}

The rest of this paper is organized as follows. Section \ref{section2} presents the problem formulation including scenario description, vehicle dynamics, energy consumption model, the intelligent driver model as well as control objectives of this paper. In Section \ref{section3}, the stochastic behaviors of human drivers are developed and the detailed design process of reinforcement learning is given. Simulation results and performance analysis are provided in Section \ref{section4}. Conclusions of this paper are presented in Section \ref{section5}.

\section{Problem formulation}  \label{section2}

\subsection {Scenario Description}

\begin{figure}
\centering
\includegraphics[width=3in]{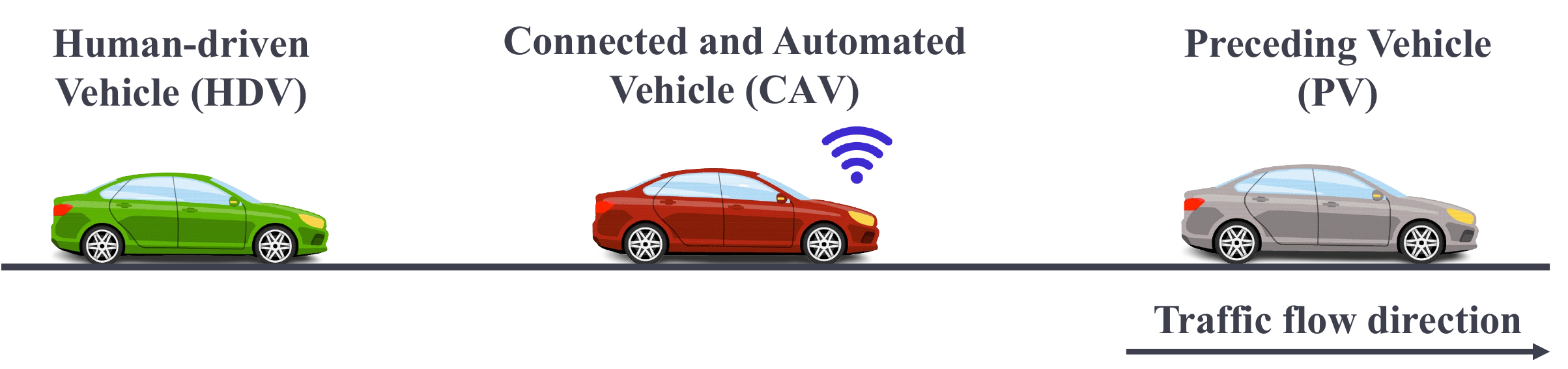}
\caption{Schematic of the 3-vehicle flow consisting of PV, CAV, and HDV.}\label{MixedTrafficSchematic}
\end{figure}

Similar to existing studies in \cite{ma2020energetic} and \cite{ozkan2021modeling}, a common scenario of mixed traffic stream is investigated in this study, where there is a human-driven preceding vehicle (PV), a CAV, and a following HDV, as shown in Fig.~\ref{MixedTrafficSchematic}. Assume the CAV can obtain velocity and gap distance information of both the PV and the following HDV through onboard sensors (e.g., millimeter-wave radars). There may exist more vehicles (CAVs or HDVs) before the PV, which means this scenario is just a fraction of a long mixed traffic flow, while the traffic before the PV is not the focus of this study. 

\subsection {Vehicle Longitudinal Dynamics}
Since this study emphasizes energy-efficient driving in a car-following scenario, only vehicle longitudinal dynamics is considered here, as described in Eq.~\eqref{eq:VehicleLongitudinalDynamics}.

\begin{equation} \label{eq:VehicleLongitudinalDynamics}
\dot v = \frac{F_{\tx d} - mgf \cos\alpha - mg \sin\alpha - 0.5C_{\tx D} A_{\tx f} \rho v^2}{\delta m}                            
\end{equation}

\noindent where $v$ is the vehicle velocity. $F_{\tx d}$ denotes the driving force of the vehicle. $m$ is vehicle mass and $g$ denotes the gravitational acceleration. $C_{\tx D}$, $A_{\tx f}$, and $\rho$ are aerodynamic drag coefficient, vehicle frontal area, and air density, respectively. $\delta$ is the rotational inertia coefficient. For simplicity, road slope is not considered here ($\alpha=0$). Table~\ref{tab:vehparams} lists the critical parameters of the longitudinal dynamics model of the vehicle.

\begin{table}
\footnotesize
\caption{Parameters of vehicle longitudinal dynamic model}
\centering
\begin{tabular}{ll}
\hline
Model Parameter & Value\\
\hline
Vehicle mass $m$ & \SI{1005}{kg}\\
Aerodynamic drag coefficient $C_{\tx{D}}$ & \SI{0.3}{}\\
Vehicle frontal area $A_{\tx{f}}$ & \SI{2.02}{m^2}\\
Rolling resistance coefficient $f$ & \SI{0.015}{}\\
Air density $\rho$ & \SI{1.206}{kg/m^3}\\
Gravitational acceleration $g$ & \SI{9.81}{m/s^2}\\
Wheel radius $r$ & \SI{0.28}{m}\\
Rotational inertia coefficient $\delta$ & \SI{1.02}{}\\
\hline
\end{tabular}
\label{tab:vehparams}
\end{table}

\subsection {Energy Consumption Model}
\begin{figure}
\centering
\includegraphics[width=3in]{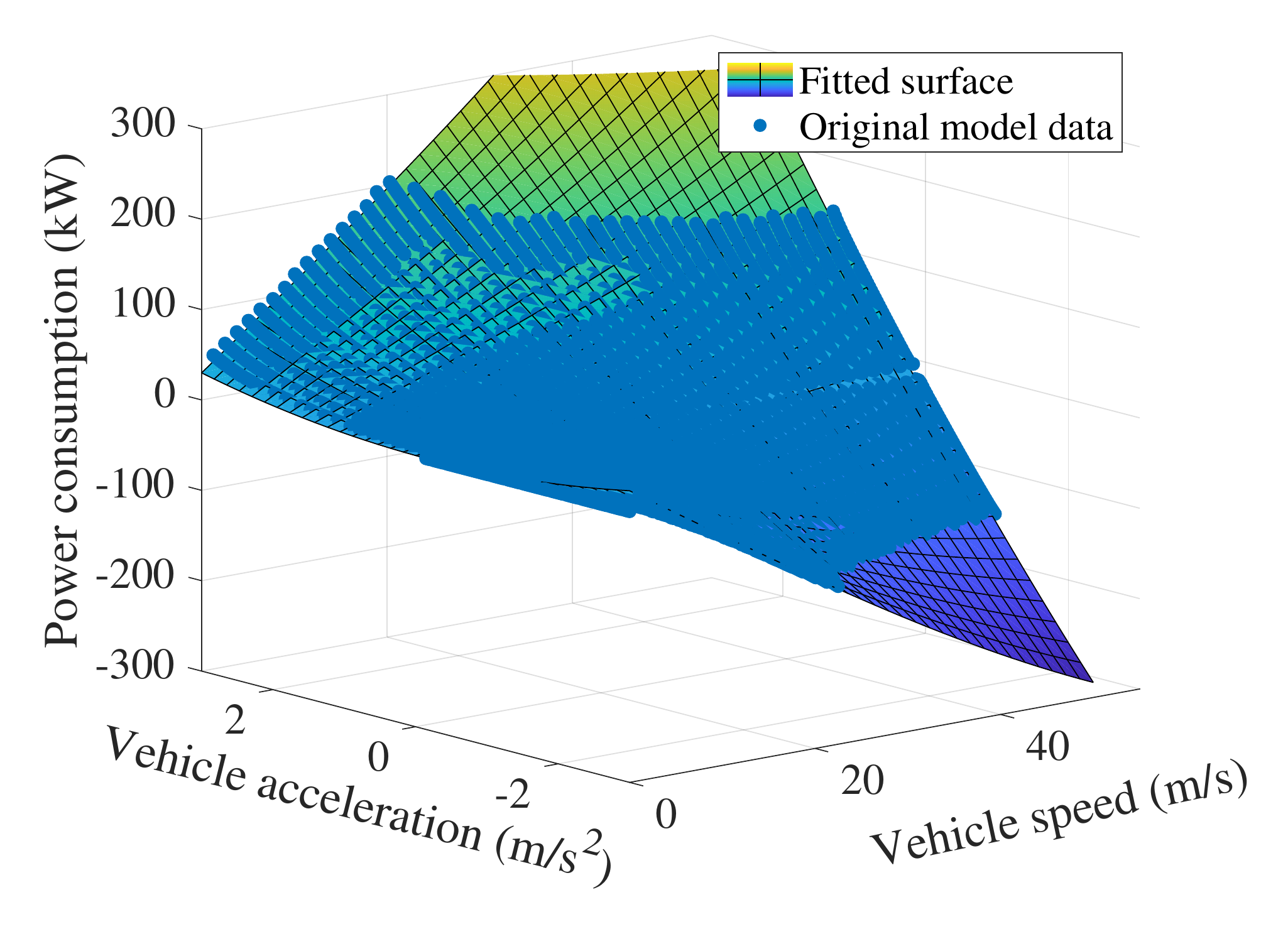}
\caption{Vehicle power consumption in terms of vehicle speed and acceleration.}\label{motorfitting}
\end{figure}

Generally, an approximated and differentiable energy consumption model in a polynomial expression is sufficient to develop the eco-driving algorithm. According to the experimental data in \cite{wang2022ecological}, the demand power of the motor $P_{\tx{mot}}$ can be written as a nonlinear function of the vehicle speed and acceleration. With acceptable accuracy, this model is free from heavy computation and complicated calibration. The detailed formation process is described in \cite{guzzella2007vehicle} and our fitted instantaneous energy consumption model is given as follows:

\vspace*{-3mm}
\begin{equation}\label{eq:ECModel}
P_{\tx{mot}}(v,a)=\sum_{i=0}^{3} \sum_{j=0}^{2} p_{ij} \, v^i(t) \, a^j(t)
\end{equation}

\noindent where $v(t)$ is the instantaneous vehicle speed; $a(t)$ is the instantaneous acceleration. $p_{ij}$ are fitting parameters. Through trial-and-error, the polynomial order and coefficients are determined with the help of MATLAB Curve Fitting Toolbox. As illustrated in Fig.~\ref{motorfitting}, a good fitting result can be obtained. The values of these coefficients are listed in Table~\ref{tab:ECModelparams}. The other coefficients, such as $p_{32}$, which are not given in Table~\ref{tab:ECModelparams}, equal to zero by default. Without loss of generality, the CAV and HDV in the mixed traffic flow are assumed to follow the same energy consumption model.

\begin{table}
\footnotesize
\caption{Parameters of energy consumption model}
\centering
\begin{tabular}{ll}
\hline
Model Parameter & Value\\
\hline
$p_{00}$ & \SI{110.3}{}\\
$p_{10}$ & \SI{422.9}{}\\
$p_{01}$ & \SI{1213}{}\\
$p_{20}$ & \SI{-0.0279}{}\\
$p_{11}$ & \SI{2484}{}\\
$p_{02}$ & \SI{2911}{}\\
$p_{30}$ & \SI{0.3557}{}\\
$p_{21}$ & \SI{1.374}{}\\
$p_{12}$ & \SI{25.19}{}\\
\hline
\end{tabular}
\label{tab:ECModelparams}
\end{table}

\subsection {Intelligent Driver Model}
The microscopic traffic model is utilized to simulate the car-following behaviors of HDVs in the transportation system. 
Inspired by the work in \cite{zhu2018modeling}, the intelligent driver model (IDM) \cite{treiber2013traffic} is used in this study for its proven performance in accident-avoidance and generating realistic acceleration profiles. In addition, the parameters in the IDM capture diverse characteristics of drivers in virtually all single-lane traffic scenarios, i.e. the car-following scenario \cite{treiber2013traffic}. 
Following is a brief description of the model:

\begin{align}
&\dot{v}=a\left(1-\left(\frac{v}{v_{0}}\right)^{\delta}-\left(\frac{s^{*}\left(v, \Delta v\right)}{s}\right)^{2}\right) 
\label{eq:IDMEquation}\\
&s^{*}(v, \Delta v)=s_{0}+\max \left(0, vT+\frac{v \Delta v}{2 \sqrt{a b}} \right)
\label{eq:desiredDistance}
\end{align}

\noindent where $v$ is the instantaneous vehicle velocity; $a$ denotes the maximum acceleration parameter; $v_{0}$ represents the desired velocity; $\delta$ is the acceleration exponent; $s^{*}$ and $s$ denote the desired gap distance and the actual gap distance to the preceding vehicle; $\Delta v$ is the speed difference to its preceding vehicle; $s_{0}$ is the minimum gap; $T$ is the time gap to the preceding vehicle; and $b$ is the comfortable deceleration parameter.


Each parameter in the IDM characterizes a specific aspect of human driving behaviors. For example, the maximum acceleration parameter $a$ represents the driver's aggressiveness in hitting the accelerator pedal. As the vehicle speed increases and approaches the desired speed $v_0$, the acceleration decreases accordingly and stabilizes around zero. The value of acceleration exponent $\delta$ determines how fast the acceleration drops when approaching the desired speed. The time gap $T$ defines the interval between the moment the rear bumper of the preceding vehicle passes an appointed location and the moment the front bumper of the following vehicle reaches the same location \cite{loulizi2019steady}. Intuitively, an aggressive driver tends to keep a relatively smaller time gap in comparison with a conservative driver. Overall, the IDM can produce smooth speed profiles and easily model various driving behaviors by tuning those parameters \cite{treiber2013traffic}.


\subsection {Control Objectives}
\begin{figure*}
\centering
\includegraphics[width=6.5in]{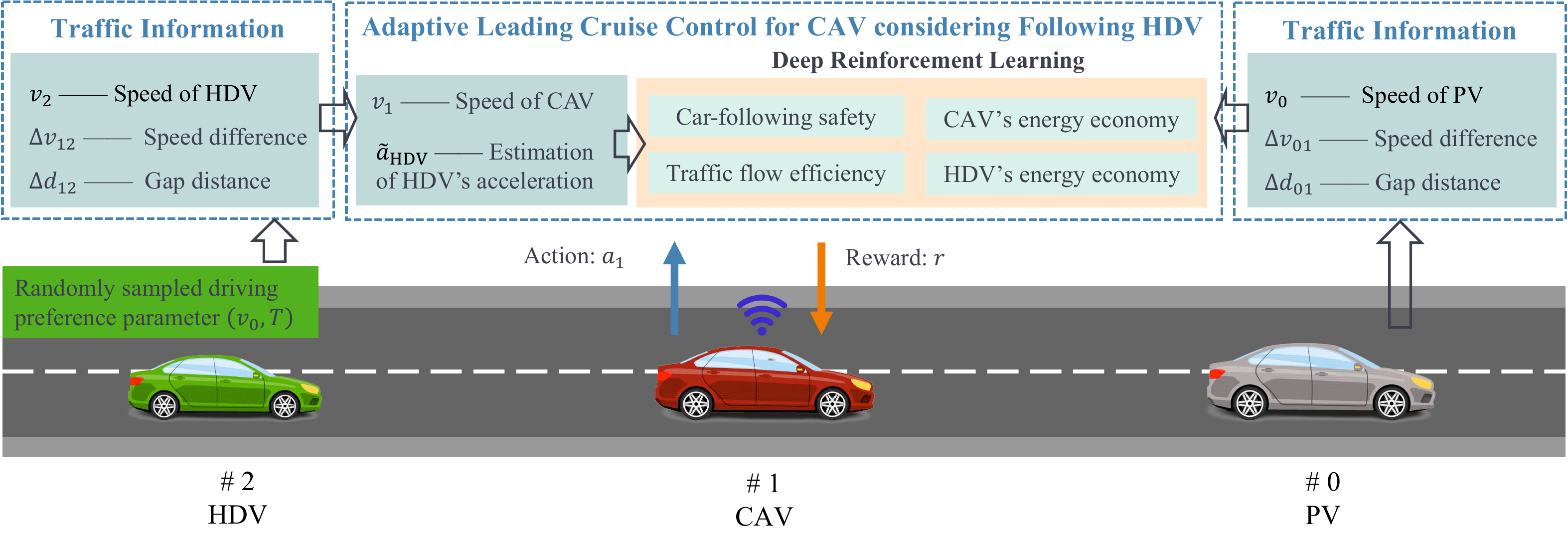}
\caption{Schematic of the leading cruise control problem addressed in this study. }\label{Schematic}
\end{figure*}

As shown in Fig.~\ref{Schematic}, in such a mixed traffic flow, only CAV is fully controllable and it has an impact on the energy performance of following HDV. The control objective of this paper is to improve the potential energy-saving benefits enabled by vehicle connectivity and autonomy through dedicated control of CAV while taking into account the diverse behaviors of the HDV. 
The control objectives are given by:


\begin{subequations} \label{eq:OCP}
\begin{align}
&a_{1}^{*}(t)=\underset{a_{1}(t)}{\arg \min }(J), t \in\left[t_{0}, t_{f}\right]\label{eq:controlVariable}\\
&\tx{with} \nonumber\\
&J\left(a_{1}(t)\right)=\int_{t_{0}}^{t_{f}} \left(P_{\tx{CAV}}(t) + P_{\tx{HDV}}(t) \right) dt \label{eq:objectiveFunction}\\
&\tx{subject\,to} \nonumber\\
& \tx{TTC}(t) > \tx{TTC}_{\min} \label{eq:TTCcons}\\
& \tx{TG}(t) < \tx{TG}_{\max} \label{eq:TGcons}\\
& a_{\min } \leq a_{1}(t) \leq a_{\max } \label{eq:controlcons}
\end{align}
\end{subequations}

\noindent where $a_1$ is the acceleration of CAV and is also the control variable of the system; $a_{1}^{*}(t)$ represents the acceleration profile with the optimal energy-saving performance within the time interval $\left[t_{0}, t_{f}\right]$. $J$ is the total cost function including both CAV's and HDV's trip energy consumption; and $P_{\tx{CAV}}$ and $P_{\tx{HDV}}$ are the instantaneous energy consumption rate of CAV and HDV as illustrated in Eq.~\eqref{eq:ECModel}. 
Besides, some constraints are imposed for collision avoidance and traffic efficiency purposes. 
Time-to-collision (TTC) indicates how much time remains before two vehicles collide and time gap (TG) here is the time headway of CAV to the preceding vehicle \cite{vogel2003comparison}.
The detailed calculation process of TTC and TG will be given in Section \ref{section3}.
In Eq.~\eqref{eq:TTCcons} and \eqref{eq:TGcons}, $\tx{TTC}_{\min}$ and $\tx{TG}_{\max}$ represent the critical values.
$a_{\min }$ and $a_{\max }$ in Eq.~\eqref{eq:controlcons} denote the lower and upper bounds of the acceleration of CAV. These constraints imply the CAV is required to follow the preceding vehicle within a moderate range. 
$\tx{TTC}_{\min}$ specifies the critical time gap to the preceding vehicle for the sake of collision and $\tx{TG}_{\max}$ prevents the spacing from being overlarge at the cost of reducing traffic throughput. The constraint on vehicle acceleration is imposed to ensure ride comfort and respect all physical limits.

\section{Adaptive leading cruise control}  \label{section3}
In this section, a RL-based adaptive leading cruise control strategy is developed. First, in order to better model the stochastic human driving behaviors, a genetic algorithm (GA) is utilized to find the optimal parameters in the IDM. The formulation of the longitudinal speed control strategy for the CAV will then be given. Different from previous research where only CAV is considered, we aim to improve the energy efficiency of both CAV and the following HDV by leveraging the predictive dynamics of HDV. The formulated optimal control problem is solved in a learning manner based on a RL framework. In the rest of this section, the detailed formulation of the reward function will be presented.

\subsection {Stochastic Driver Behaviors}
To model and calibrate the realistic human driving behaviors, the field-test vehicle trajectory data from the NGSIM project is utilized \cite{alexiadis2004next}. The trajectory data were collected on eastbound I-80 in Emeryville, California using an array of synchronized digital video cameras and then transcribed into the format of vehicle trajectory with a customized software named NGVIDEO. The original dataset encompasses abundant information such as a buildup of traffic congestion, semi-congested state, and full congestion during peak hours. A car-following extraction filter proposed in \cite{wang2017capturing} was applied to extract the car-following episodes. A total of 923 car-following episodes at a sampling rate of 10 Hz were extracted and used in our study.
The car-following speed information of different human drivers in the NGSIM dataset is applied to calibrate the IDM, as introduced in Section \ref{section2}.
Aiming at minimizing the difference between the actual observed speed in the dataset and the simulated speed of the IDM, the objective function is defined as minimizing the root mean square percentage error (RMSPE) of speed difference:

\begin{equation}\label{eq:GAObjectiveFunction}
\tx{min} \sqrt{\frac{\sum_{i=1}^{N} \left(v_i^{\tx{sim}} - v_i^{\tx{obs}}\right)^2}{\sum_{i=1}^{N} \left(v_i^{\tx{obs}}\right)^2}}
\end{equation}
where $N$ denotes the total number of car-following episodes ($N=923$); $v_i^{\tx{obs}}$ is the actual speed of the $i$th car-following episode; and $v_i^{\tx{sim}}$ is the simulated speed of the IDM.

The driving preference parameters in the IDM are optimized using GA, where the process is given as follows:
\begin{enumerate}
    \item Initialization is carried out on a population of $m$ individuals and each individual represents one driving preference parameter in the IDM;
    \item A predefined objective function \eqref{eq:GAObjectiveFunction} is used to determine the fitness of each individual within the population;
    \item Crossover and mutation are implemented to produce the children generation from the parents' generation;
    \item Algorithm terminates when the stopping criteria are met.
\end{enumerate}

\begin{figure}
\centering
\includegraphics[width=3.3in]{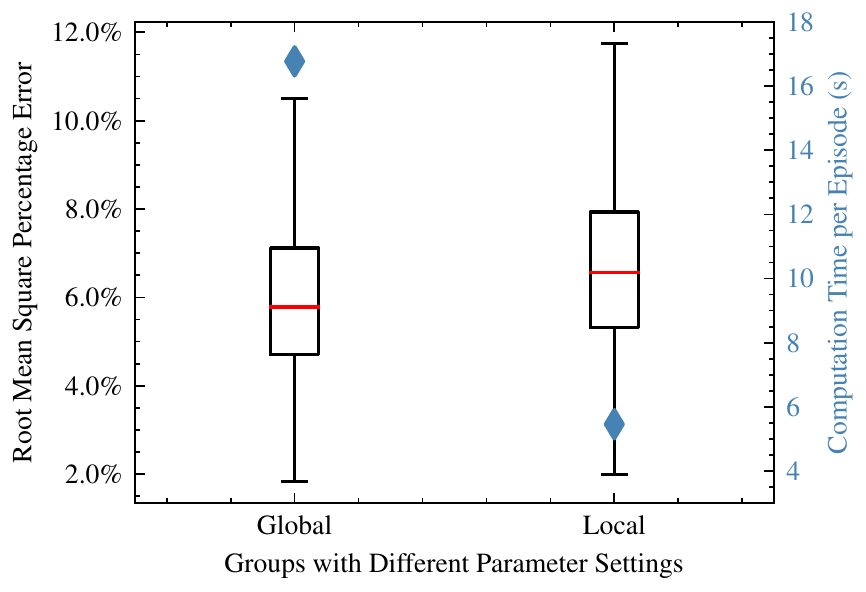}
\caption{Distributions of the root mean square percentage error under two groups of parameter settings.}\label{CarFollowingFittingError}
\end{figure}

In order to balance the optimization performance and computational efficiency, instead of directly optimizing all parameters $\left (a, v_0, s_0, T, b \right)$ in the IDM, only two parameters $\left (v_0, T \right)$ are optimized while the others are set as constant after trial-and-error. According to simulation results, the distributions of $\left (v_0, T \right)$ share a similar pattern with that of the maximum driving speed of human drivers in the NGSIM dataset, while the other parameters cluster around boundary values, which means $\left (v_0, T \right)$ can better reflect the styles of human drivers. As shown in Fig.~\ref{CarFollowingFittingError}, the boxplot gives the distribution of the RMSPE under two groups with different parameter settings. The global group means all the five driving preference parameters are optimized simultaneously, while in the local group, only $v_0$ and $T$ are optimized with other parameters set as constants. The blue diamond symbols indicate the average computation time for each episode.
With acceptable sacrifice in optimization performance (mean value of RMSPE increasing from \SI{6.0}{\%} to \SI{6.7}{\%}), the average computation time drops from \SI{16.8}{s} to \SI{5.5}{s} for each optimization process, which would greatly improve the training efficiency for RL. 

\begin{figure}
\centering
\includegraphics[width=3.3in]{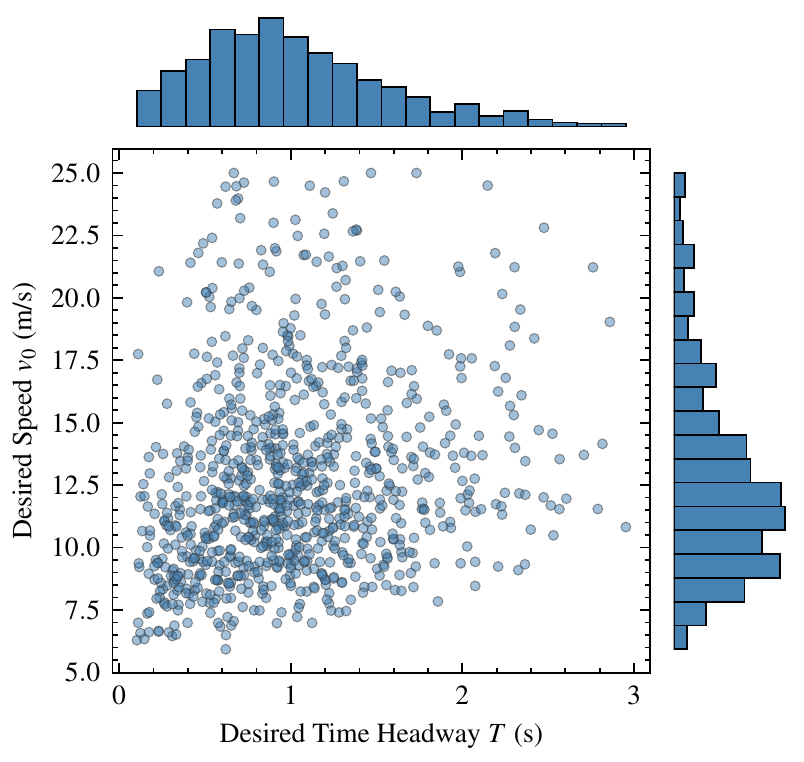}
\caption{Distributions of $v_0$ and $T$.}\label{JointDistribution}
\end{figure}

Fig.~\ref{JointDistribution} gives the distributions of $v_0$ and $T$ after optimization. 
The correlation coefficient of the two variables equals $0.24$, which means they are not independent, but follow a joint distribution. Each circle in Fig.~\ref{JointDistribution} represents the specific preferences of the corresponding human driver. Random sampling from this joint distribution is used to replicate the different preferences of real human drivers and the uncertainty in real traffic situations.


\subsection {Preliminaries of Reinforcement Learning} 
This subsection introduces the fundamentals of RL. The modeling basis of RL is the markov decision process, where an agent interacts with a stochastic environment \cite{sutton2018reinforcement}. At each time step $t$, the agent observes the state $s_t$ from its surrounding environment and then selects an action $a_t$ to execute based on the current policy $\pi (a_t|s_t)$. The environment returns a scalar reward $r_t$ to the agent and transits to a new state $s_{t+1}$. The goal of the agent is to obtain a policy that maximizes the discounted total reward. The optimal policy can be written as:


\begin{equation}\label{eq:optimalPolicy}
\pi^* = \underset{\pi}{\arg \max } \mathbb{E}\left(\sum_{t=0}^{T} \gamma^{t} r\left(s_{t}, a_{t}\right)\right)
\end{equation}

\noindent where $\gamma \in (0,1]$ denotes the discount factor. Combined with deep learning, deep reinforcement learning (DRL) uses neural networks to approximate the Q-value function \cite{mnih2015human}, policy $\pi (a_t|s_t)$ \cite{lillicrap2015continuous} or system model. Since the selected control variable (i.e., acceleration of CAV) is continuous, the deep deterministic policy gradient (DDPG) algorithm is utilized for its capability of dealing with continuous action spaces  \cite{lillicrap2015continuous}. The policy-network, parameterized by $\theta^\mu$, is used to deterministically map the state into the action as $a_t = \pi(s_t|\theta^\mu)$. The Q-value network, parameterized by $\theta^Q$, is used for value function approximation. The Q-value network is updated by minimizing the loss function as follows:


\begin{equation}
L=\frac{1}{N} \sum_{i}\left(y_{i}-Q\left(s_{i}, a_{i} \mid \theta^{Q}\right)\right)^{2}
\end{equation}

\noindent where $y_{i}=r_{i}+\gamma Q^{\prime}\left(s_{i+1}, \mu^{\prime}\left(s_{i+1} \mid \theta^{\mu^{\prime}}\right) \mid \theta^ {Q^{\prime}}\right)$. Then the gradients $\nabla_{a} Q(s, a)$, calculated by the Q-value network, are passed to the policy-network to update the weights and biases using stochastic gradient descent (SGD):

\begin{equation}
\nabla_{\theta^{\mu}} J=\frac{1}{N} \sum_{i} \nabla_{a} Q\left(s, a \mid \theta^{Q} \right)  \nabla_{\theta^{\mu}} \mu \left(s \mid \theta^{\mu} \right)
\end{equation}



\subsection {Design of Reinforcement Learning Algorithm} 
As inputs into the agent, the state space should provide enough information about the current traffic situation. The speed of three vehicles, their relative speed and the relative gap distance concurrently make up the state space, which can be expressed as $\textit{S} = \{v_0, v_1, v_2, \Delta v_{01}, \Delta v_{12}, \Delta d_{01}, \Delta d_{12} \}$. The control input, i.e., CAV's acceleration, makes up the action space, which is written as: $\textit{A} = \{a_1\}$. The physical meaning of these variables can be found in Fig.~\ref{Schematic}. The objective of this study is to design an adaptive leading cruise control strategy to regulate the CAV and achieve four sub-objectives: car-following safety, traffic efficiency, energy economy of CAV, and energy economy of HDV. 
By incorporating the energy economy of the following HDV into the reward function for the first time, we aim to improve the overall energy efficiency from a statistical perspective. The reward function capturing the four sub-objectives, as listed below, is adopted.

\begin{itemize}[wide]
    \item [1)]
    \textit{Car-following Safety:} The highest priority should always be given to safety to avoid collisions. TTC, which indicates how much time remains before collision, is treated as the metric for safety, as given by:
    \begin{equation}
    \tx{TTC}(t) = -\frac{\Delta d_{01}(t)}{\Delta v_{01}(t)}
    \end{equation}
    where $0$ is the index for preceding vehicle and $1$ is the index for CAV; $\Delta d_{01}$ denotes their gap distance and $\Delta v_{01}$ represents their relative speed $(\Delta v_{01} = v_0 - v_1)$. A smaller TTC value is associated with higher crash risk, and vice versa \cite{vogel2003comparison}. The lower bound of TTC, as a critical boundary differentiating `safe driving' and `dangerous driving', should be clearly stated. A final value of $4s$ is selected and if TTC is smaller than $4s$, the corresponding negative reward will be given \cite{zhu2020safe}. Intuitively, a smaller TTC value comes with a bigger penalty. Therefore, the reward corresponding to car-following safety is given as follows:
    \begin{equation}
    R_{\tx{safe}} = \begin{cases}\log \left(\frac{\tx{TTC}}{4}\right) & 0 \leq \tx{TTC} \leq 4 \\ 0 & \text { otherwise }\end{cases}
    \end{equation}
    \item [2)]
    \textit{Traffic Efficiency:} High traffic efficiency defined in this study indicates a small and safe TG from the CAV to the preceding vehicle. A small TG within the safe range usually means a higher roadway capacity \cite{zhang2007examining}. There are slight differences regarding the legal or recommended safety distances in different countries. For instance, several driver training programs in the United States claim that it is hard for human drivers to safely follow a preceding car within a time gap of less than $2s$ \cite{michael2000headway}. In Germany, a headway of larger than $1.8s$ is recommended. A time gap of $3s$ is recommended by the Swedish National Road Administration in rural areas \cite{vogel2003comparison}.
    In this study, in order not to sacrifice the potential in energy saving, only an overlarge time gap will be penalized and its boundary value is set as $2.5s$. The reward in terms of traffic efficiency is constructed as follows:
    \begin{equation}
    \tx{TG}(t) = \frac{\Delta d_{01}(t)}{v_1(t)}
    \end{equation}
    
    \begin{equation}
    R_{\tx{eff}} = \begin{cases} -1 & \tx{TG} \geq 2.5 \\ 0 & \text { otherwise }\end{cases}
    \end{equation}
    \item [3)]
    \textit{Energy Economy of CAV:} One major objective of eco-driving is to save as much energy as possible. We aim to find an optimal speed trajectory leading to the minimum energy consumption by controlling vehicle acceleration. Using the instantaneous energy consumption model introduced in Section \ref{section2}, the immediate energy consumption in each time step can be estimated. The reward in terms of the energy economy of CAV can be described as:
    \begin{equation}
    R_{\tx{CAV}} = - \frac{P_{\tx{CAV}}}{20000} \times \Delta t
    \end{equation}
    where the constant value of 20000 is used to scale the reward value into the range of $\left[ -1, 0 \right]$, and $\Delta t$ is time step and equals to $0.1s$.
    \item [4)]
    \textit{Energy Economy of HDV:} Since the HDV is not controllable, we can only improve its energy economy by leveraging the interaction between the HDV and the regulated CAV. However, it is challenging to precisely predict how the HDV reacts to the surrounding traffic due to the uncertain and diverse human driving behaviors. To this end, the HDV behavior will be learned and predicted based on the data collected in real time. Specifically, at the initial stage (first $5s$) of car-following, the expected HDV acceleration is initialized based on the average traffic situation. As the car-following process of HDV continues, the historical speed information can be used to learn and fit the two driving preference parameters $\left (\tilde v_0, \tilde T \right)$ in the IDM. Besides, with real-time access to the gap distance $\Delta d_{12}$ between CAV and HDV, the expected acceleration of the HDV can be derived in \eqref{eq:HDVAcc}.
    

    \begin{equation}\label{eq:HDVAcc}
    \tilde a_{\tx{HDV}} = \begin{cases} \mathbb{E}\left(\sum_{i=1}^{N} f_{\tx{IDM}} \left(v_{0_i}, T_i \right) \right) & 0 \leq t < 5s \\ f_{\tx{IDM}} \left( \tilde v_0, \tilde T \right) & t \geq 5s  \end{cases}
    \end{equation}
    
    \noindent where $f_{\tx{IDM}}$ is the dynamics of IDM as described in Eqs.~\eqref{eq:IDMEquation} and \eqref{eq:desiredDistance}. 
    With this estimated acceleration $\tilde a_{\tx{HDV}}$, as well as the HDV speed $v_{\tx{HDV}}$ which can be obtained from on-board sensors, the reward in terms of the energy economy of of HDV can be calculated as follows: 
    
    \begin{equation}\label{eq:HDVReward}
    R_{\tx{HDV}} = - \frac{\tilde P_{\tx{HDV}}}{20000} \times 0.1 = - \frac{P_{\tx{HDV}} (v_{\tx{HDV}}, \tilde a_{\tx{HDV}})}{20000} \times \Delta t
    \end{equation}
    
\end{itemize}

Finally, the reward function is constructed based on a linear combination of all aforementioned features.

\begin{equation}
r = R_{\tx{safe}} + R_{\tx{eff}} + R_{\tx{CAV}} + R_{\tx{HDV}}
\end{equation}

\section{Simulation results}  \label{section4}

\subsection{Simulation Setup}
The speed profile of the preceding vehicle is also sampled from the NGSIM dataset to represent the real-world scenario, and it lasts for $30s$ with the sampling rate being $0.1s$. Correspondingly, there are 300 steps in each training episode and each step equals $0.1s$ in real life.
All the state variables are fed into the policy network $s_t = \{v_0(t), v_1(t), v_2(t), \Delta v_{01}(t), \Delta v_{12}(t), \Delta d_{01}(t), \Delta d_{12}(t) \}$ and it outputs the acceleration of CAV directly $a_t = a_1 (t)$. The inputs of the Q-value network are the state and action and it outputs a scalar Q-value $Q(s_t, a_t)$.

A variety of DDPG models were trained in this stochastic mixed traffic environment to find out the suitable network structure. Finally, a structure of three fully connected hidden layers (200-100-50 neurons in each layer) with rectified linear unit (ReLU) activations is adopted. The output layer of the policy network is activated by a \textit{tanh} function to map into range $\left[-1,1 \right]$. After that, a linear mapping is applied to transform the outputted acceleration between $\left[-3,3 \right] m/s^2$. An overview of the hyperparameters is given in Table~\ref{tab:hyperparameters}. Besides, in order to compensate for the optimization error in Fig.~\ref{CarFollowingFittingError} and better consider the randomness in human driving behaviors, a random noise $\xi$ is added to the outputted acceleration of HDV. A kinematic point-mass model is used to update the environment state:

\begin{subequations} \label{eq:kinematicPoint-massModel}
\begin{align}
& v_1 (t+1) = v_1 (t) + a_1 (t) \times \Delta t \\
& v_2 (t+1) = v_2 (t) + \left( a_2 (t) \times \left(1+\xi\right) \right) \times \Delta t    \\
& \Delta v_{12} (t+1) = v_1 (t+1) - v_2 (t+1) \\
& \Delta d_{12} (t+1) = \Delta d_{12} (t) + \frac{\Delta v_{12} (t)+\Delta v_{12} (t+1)}{2} \times \Delta t
\end{align}
\end{subequations}

\noindent where $\xi$ is the random noise ranging from $0\%$ to $5\%$.

\begin{table}
\footnotesize
\caption{Parameter settings}
\centering
\begin{tabular}{lll}
\hline
Hyperparameter & Value & Description\\
\hline
$\gamma$ & 0.9 & Discount factor\\
\textit{N} & 1024 & Number of samples for SGD update\\
\textit{D} & 20000 & Number of samples in replay buffer\\
Lr\_A & 0.001 & Learning rate for policy network\\
Lr\_C & 0.001 & Learning rate for Q-value network\\
\hline
\end{tabular}
\label{tab:hyperparameters}
\end{table}

\subsection{Reference Control Algorithm}
For the purpose of comparison, a reference strategy that did not consider the following HDV in the optimization was also developed. The objectives of this reference strategy including car-following safety, traffic efficiency, and the energy economy of CAV, share the same structure as illustrated in Section \ref{section3}. Specifically, the state space in this reference strategy is $\textit{S} = \{v_0, v_1, \Delta v_{01}, \Delta d_{01}\}$, the action space is defined as $\textit{A} = \{a_1\}$. The reward function can be described as follows:

\begin{equation}
r_{\tx{ref}} = R_{\tx{safe}} + R_{\tx{eff}} + R_{\tx{CAV}}
\end{equation}

\subsection{Training Results}
\begin{figure}
\centering
\includegraphics[width=3in]{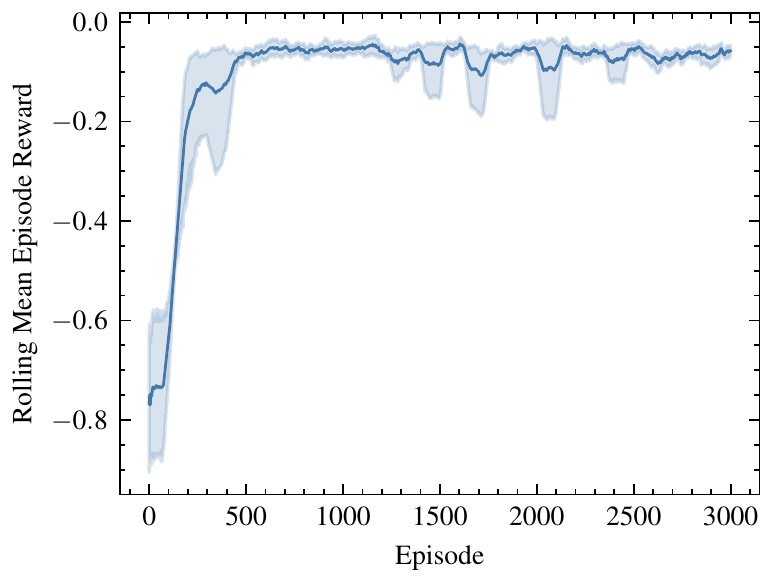}
\caption{Training result of DDPG.}\label{6_trainingResults}
\end{figure}

Considering that stochasticity significantly affects training performance, the training was repeated eight times with different random seeds. Each training consists of 3000 episodes. At the beginning of each episode, the driving preference set $(v_0, T)$ of HDV is randomly sampled from the joint distribution as shown in Fig.~\ref{JointDistribution}. This is to replicate the driving diversity of human drivers in real traffic.  Fig.~\ref{6_trainingResults} gives the trajectory of rolling mean episode reward with respect to the training process. The blue solid line is the rolling mean episode reward and the translucent area indicates a 95\% confidence interval over eight runs. A steep increase can be observed within the first 500 episodes. As the training progresses, the reward trajectory starts to converge, thereby proving that the designed reward function is stable and effective.

\begin{figure}
\centering
\includegraphics[width=3in]{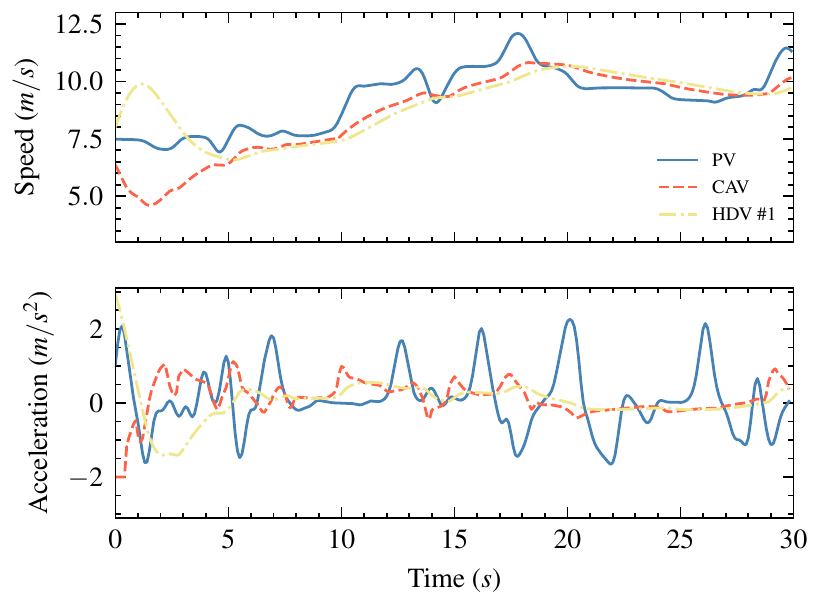}
\caption{A random sample of car-following behaviors.}\label{7_Speed_acceleration_profiles}
\end{figure}

\begin{figure}[b]
\centering
\includegraphics[width=3in]{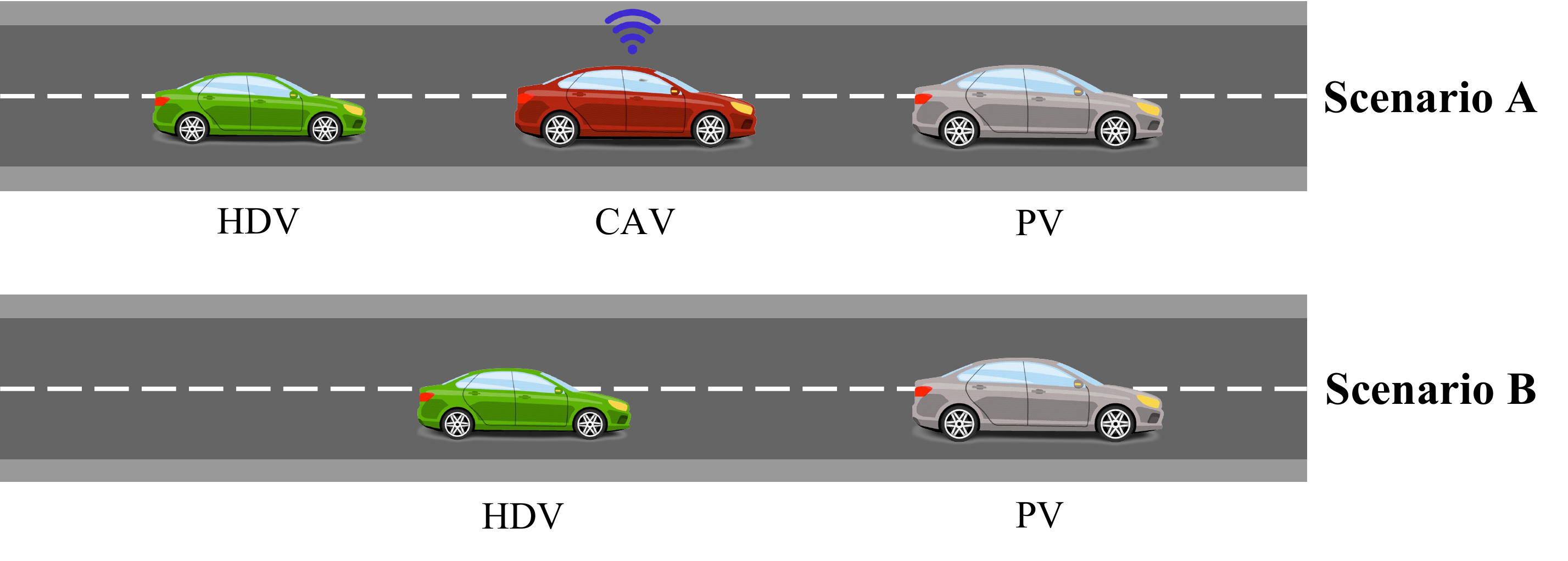}
\caption{Illustration of two driving scenarios.}\label{DrivingScenario}
\end{figure}

As shown in Fig.~\ref{7_Speed_acceleration_profiles}, a randomly sampled car-following episode simulated by DDPG is given and shows the speed and acceleration of the three vehicles in the mixed traffic flow. It is obvious that the acceleration amplitude of CAV is much smaller than that of the preceding vehicle. As a result, maneuvers like drastic acceleration and hard braking can be avoided for both CAV and HDV by adopting the proposed leading cruise control algorithm. To investigate the energy economy performance of the proposed algorithm, we compare the trip energy consumption of HDV under two scenarios, as shown in Fig.~\ref{DrivingScenario}. In scenario A, the CAV drives after the human-driven PV and the HDV follows the CAV, where the CAV is actively regulated to lead the HDV. In scenario B, the HDV directly follows after the PV. 

\begin{figure}
\centering
\includegraphics[width=3.3in]{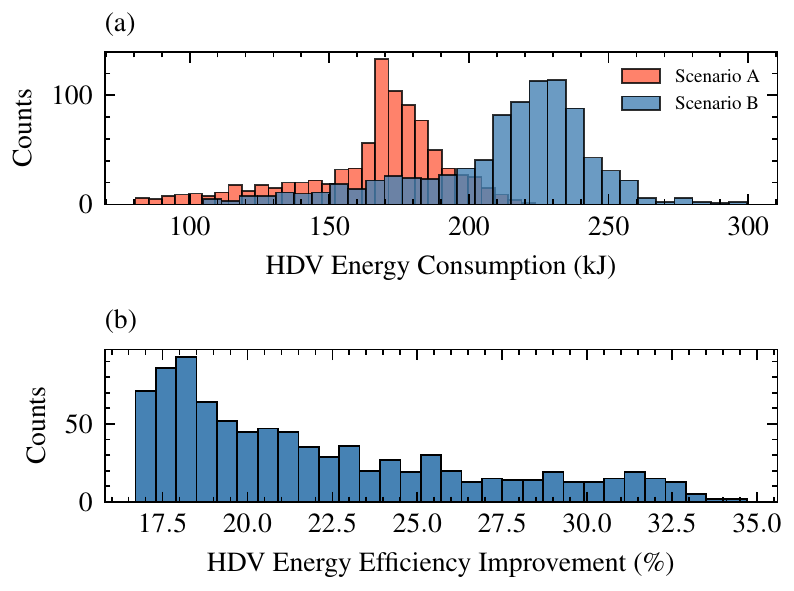}
\caption{Energy consumption of HDV under two driving scenarios.}\label{results_1}
\end{figure}

As shown in Fig.~\ref{results_1}, the trip energy consumption of all the 923 HDVs under two different driving scenarios (i.e., w/o CAV) is compared. The \textit{x}-axis shows how much energy is consumed and the \textit{y}-axis indicates the counts, i.e., the frequency. It can be observed that the columns in Scenario A move towards the left when compared with those in Scenario B. The average energy consumption of HDV is \SI{165.38}{kJ} for Scenario A and \SI{212.39}{kJ} for Scenario B, i.e., a reduction of average energy consumption by up to \SI{22.1}{\%}. 
The lower part in Fig.~\ref{results_1} presents a point-to-point comparison of energy efficiency improvement in percentage, with a minimum improvement of \SI{16.7}{\%} and a maximum improvement of \SI{34.7}{\%}. It can be interpreted that the HDVs with diverse characteristics can all benefit from the adaptive leading cruise control. The CAV in this traffic flow can be regarded as a virtual controller that optimizes the driving decisions of the CAV to positively interact with the following HDV. As shown in Fig.~\ref{7_Speed_acceleration_profiles}, CAV has smoother speed and acceleration profiles than those of PV and leads the HDV to behave similarly, i.e., eco-driving manner.

\begin{figure}
\centering
\includegraphics[width=3.3in]{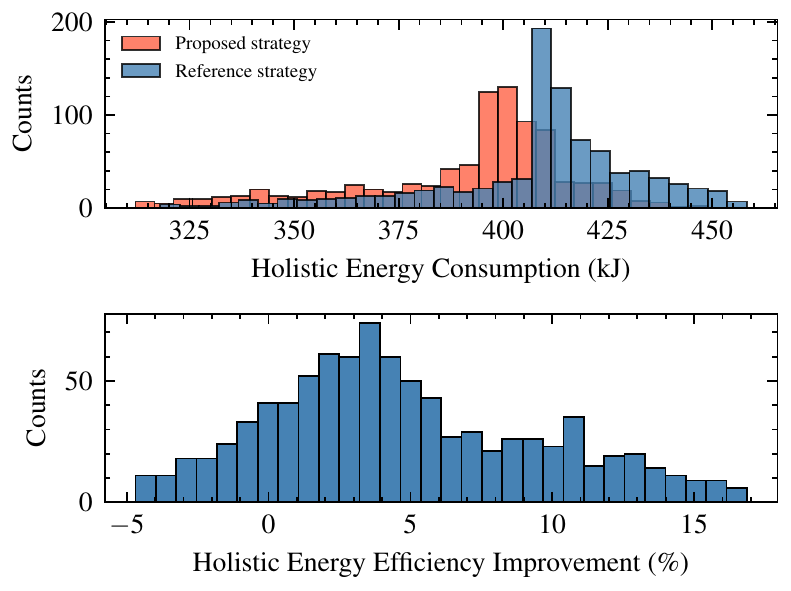}
\caption{Holistic energy consumption under two strategies.}\label{results_2}
\end{figure}

In addition, the impact of CAVs on the following HDV regarding energy efficiency is further justified by quantitatively comparing the proposed strategy (with consideration of HDV in the optimization) and the reference strategy (without consideration of HDV in the optimization) when uncertain and diverse human driving behaviors are present. The total energy consumption for comparison is calculated as follows:


\begin{equation}
\left \{ \sum_{i=1}^{2} \int_{t_{0}}^{t_{f}} P_{i} dt \right \}_j, \quad j=1,2,...,N
\end{equation}

\noindent where $i=1$ refers to CAV and $i=2$ refers to HDV, as depicted in Fig.~\ref{Schematic}. $N$ is the number of HDVs (923 in this study). 
As shown in Fig.~\ref{results_2}, by taking into account the energy economy of HDV in the proposed strategy, the total energy consumption of two vehicles is remarkably reduced for most scenarios with diverse human driving behaviors, when compared to the reference strategy. A similar analysis of point-to-point percentage improvement is also conducted and shown in the lower part of Fig.~\ref{results_2}. The proposed strategy achieves positive improvement in most scenarios. With the proposed adaptive leading cruise control strategy, the holistic energy efficiency can be increased by up to 14.43\% and the average improvement is 4.38\%. 
However, it is worth noting that there is still a small portion of cases (138 out of 923 drivers, about 15\%) with more energy consumed by two vehicles when compared to the reference strategy. The causes of negative optimization mainly come from the uncertainties in HDV, i.e., manually added noise in the acceleration and the corresponding prediction error of HDV dynamics, leading to non-deterministic optimization performance.
Note that the proposed adaptive leading cruise control strategy can statistically improve the energy efficiency (i.e., an expectation of positive results) and most HDVs will benefit from it.

\begin{figure}
\centering
\includegraphics[width=3.3in]{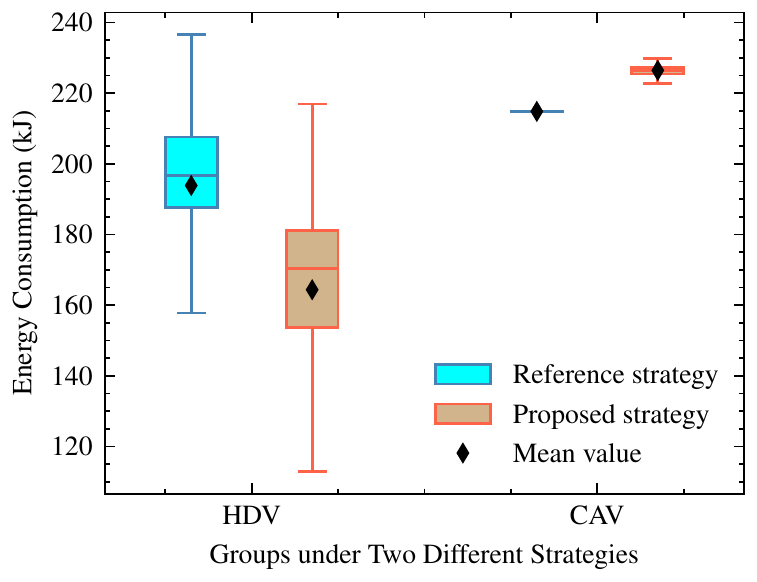}
\caption{Corresponding comparison of energy consumption of HDV and CAV under two strategies.}\label{11_EnergyConsumption_HDV_CAV}
\end{figure}

Fig.~\ref{11_EnergyConsumption_HDV_CAV} further gives the distributions of energy consumption of HDV and CAV under two strategies. In the reference strategy, the energy consumption of CAV consistently equals \SI{214.85}{kJ} following the same PV. In contrast, in the proposed strategy, with a slight sacrifice in the energy economy of CAV, the average energy consumption of HDV drops from \SI{193.89}{kJ} to \SI{164.38}{kJ} (i.e., 15.22\% reduction). 
Therefore, it is further demonstrated that by considering the HDV in the reward function, the overall energy efficiency can be improved from a statistical perspective, despite of the diverse driving preferences of HDVs. The numerical summary is given in Table~\ref{tab:ECComparison}.

\begin{table*}[htbp]
\footnotesize
\caption{Energy consumption of the mixed traffic flow under diverse driving characteristics}
\centering
\begin{tabular}{@{}cccccccc@{}}
\toprule
\multirow{2}{*}{}                                           & \multicolumn{3}{c}{Proposed Strategy}                                                                                                                                                                        & \multicolumn{3}{c}{Reference Strategy}                                                                                                                                                                       & \begin{tabular}[c]{@{}c@{}}Energy Efficiency\\ Improvement\end{tabular} \\ \cmidrule(l){2-8} 
                                                            & \begin{tabular}[c]{@{}c@{}}HDV's Energy\\ Consumption\end{tabular} & \begin{tabular}[c]{@{}c@{}}CAV's Energy\\ Consumption\end{tabular} & \begin{tabular}[c]{@{}c@{}}Total Energy\\ Consumption\end{tabular} & \begin{tabular}[c]{@{}c@{}}HDV's Energy\\ Consumption\end{tabular} & \begin{tabular}[c]{@{}c@{}}CAV's Energy\\ Consumption\end{tabular} & \begin{tabular}[c]{@{}c@{}}Total Energy\\ Consumption\end{tabular} &                                                                         \\ \midrule
\begin{tabular}[c]{@{}c@{}}Most\\ Improvement\end{tabular}  & 135.59 kJ & 221.03 kJ & 356.62 kJ & 201.91 kJ & 214.85 kJ & 416.76 kJ         & 14.43\%     \\
\begin{tabular}[c]{@{}c@{}}Least\\ Improvement\end{tabular} & 210.81 kJ & 227.11 kJ & 437.92 kJ & 202.52 kJ & 214.85 kJ & 417.37 kJ         & -4.92\%     \\
\begin{tabular}[c]{@{}c@{}}Mean\\ Improvement\end{tabular}  & 164.38 kJ & 226.46 kJ & 390.84 kJ & 193.89 kJ & 214.85 kJ & 408.74 kJ         & 4.38\%   \\ \bottomrule
\end{tabular}
\label{tab:ECComparison}
\end{table*}

\subsection{Generalization Test}
\begin{figure}
\centering
\includegraphics[width=3.3in]{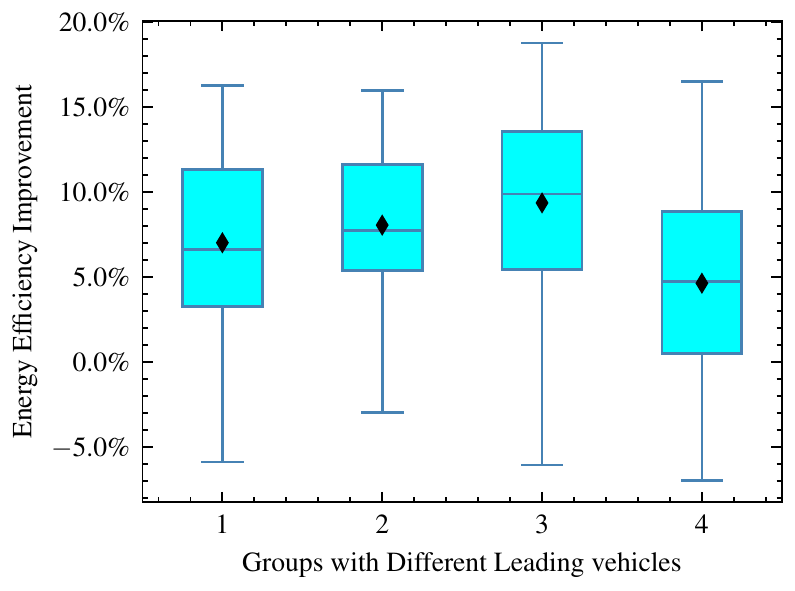}
\caption{Corresponding comparison of energy consumption of HDV and CAV under two strategies.}\label{12_Generalization}
\end{figure}

To evaluate the generalization capability of the adaptive leading cruise control strategy, four different speed profiles for the PV are randomly sampled from the NGSIM dataset. The same Monte Carlo analyses are performed to reveal the energy-saving potential. 
As shown in Fig.~\ref{12_Generalization}, similar conclusions can be obtained that the proposed strategy helps improve the overall energy efficiency in most cases with average improvements of 7.00\%, 8.04\%, 9.35\% and 4.62\% for the four cases. Simulation results demonstrate a good generalization capability of the proposed control algorithm under different driving scenarios.

\section{Conclusion}   \label{section5}
This study proposes an adaptive leading cruise control strategy for CAV while considering the diverse characteristics of following HDV in a mixed traffic flow scenario. The diversity of human drivers' behaviors is modeled using the IDM and the NGSIM dataset is used to calibrate the IDM to reflect a realistic distribution of human driving preferences. In addition to the objectives including car-following safety, traffic efficiency, and the energy economy of CAV, the energy economy of the following HDV is also incorporated into the reward function design, aiming at improving the overall energy efficiency of this mixed traffic flow. Historical speed information of HDV is utilized to learn its driving patterns and predict its acceleration of the next step. Through a Monte Carlo simulation, the proposed adaptive leading cruise control strategy exhibits an average improvement of 4.38\% in terms of traffic energy efficiency under the uncertain driving environment with diverse human driving behaviors. With a little sacrifice in the energy economy of CAV, it can help the following HDV save energy substantially. Moreover, similar statistically positive results can be observed during testing scenarios.


\ifCLASSOPTIONcaptionsoff
  \newpage
\fi

\bibliographystyle{IEEEtran}
\bibliography{refs}

\end{document}